\documentclass[11pt]{article}
\usepackage{moriond,epsfig}

\def\be{\begin{equation}}
\def\ee{\end{equation}}
\def\bea{\begin{eqnarray}}
\def\eea{\end{eqnarray}}

\begin{document}


\vspace*{4cm}
\begin{center}
{\Large\bf SEARCH FOR NEW PARTICLES AT CDF II}\\
\vspace{1cm}
{\large Simona Rolli}\\
\vspace{0.3cm}
{\it Department of Physics \& Astronomy , TUFTS University,\\
Medford, MA, 02155 USA}\\
{\it E-mail: rolli@fnal.gov}
\end{center}

\begin{figure}[ht]
\begin{center}
\psfig{figure=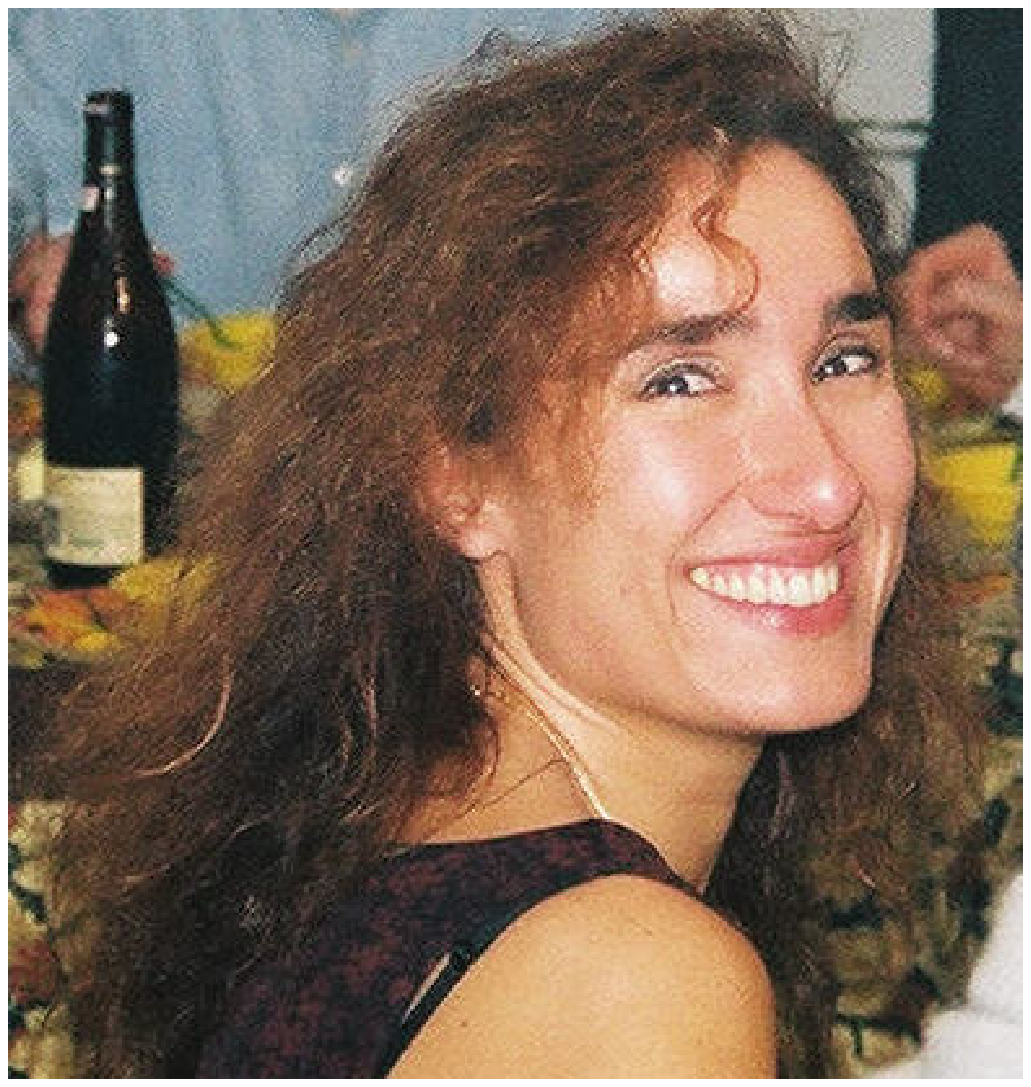,height=1.5in}
\end{center}
\end{figure}

\begin{center}
We report on the first results from the CDF experiment
on search for physics beyond the Standard Model using data from the
upgraded TeVatron collider running $p\bar p$ collisions at $\sqrt{s}$ = 1960 GeV.
These first results, although obtained with a total integrated luminosity lower than the 
total integrated luminosity collected during Run I, are already competitive if not better, 
due to the increase in the center of mass energy and the improved detector capability.
\end{center}

\section{\bf{Introduction: Run II collider and detector upgrades and search strategies at CDFII}}

\subsection{ \bf{TeVatron upgrades}}
The run I data taking period at the TeVatron ended in February 1996. Since 
then the collider and both the detectors ( CDF and D0) underwent substantial
upgrades. \par
The energy of the beams has been increased from 900 GeV to 980 GeV. 
A new syncrothron (`` main injector'') has been built in a new tunnel. 
The main injector
together with a debuncher-accumulator-recycler complex allows for 
faster production of 
antiprotons and the possibility of reusing them after they are rescued 
in the recycler.
In run I the luminosity reached 1.5$\times$10$^{31}$cm$^{-2}$sec$^{-1}$ 
and was 
obtained with a 6 on 6 proton-antiproton bunches in the collider with 
an interbunch 
time of 3.5${\mu}$sec. The luminosity ultimately  planned for run II 
is 3.3$\times 10^{32}$cm$^{-2}$sec$^{-1}$
and it  will be obtained with 36 on 36 proton-antiproton bunches with 
interbunch time of 396 ns.
So far, about 120 pb$^-1$ of integrated luminosity has been delivered and CDF has used between 
50 and 90 pb$^-1$ for its first physics results.

\subsection{\bf{CDF Detector upgrades}}
Many components of the CDF detector have been replaced or improved with 
respect to Run I\cite{tdr}.
Only the central calorimeter, the solenoid and part of the muon system
have been retained, although in general the electronics has been upgraded to
cope with the new interaction time.
Going from the inside to the outside of the detector, CDF has a new 
tracking system, composed of several silicon detectors 
( L00, SVXII and ISL) and a drift chamber.
This improved tracking system allows for particle detection extremely close to the beam pipe ( the inner most 
silicon layer sits at 2.5 cm from the beam pipe) and
extends the coverage in $\eta$ up to 2 ( it was 1.4 in Run I). 
The $z$ coverage has increased to cover the full luminous region. 
3-D track reconstruction  is possible with impact parameter resolution $\sigma_{phi} < 30~  \mu$m
and $\sigma_z < ~60 \mu$m.
\par
A completely new open cell drift chamber (Central Outer Tracker ) with maximum 
drift time of 100 ns allows for better stereo capabilities
in tracking reconstruction in respect to Run I ($\Delta p_T/p_T < 0.001$). 
It also provides $dE/dx$ information. 
\par
Between the COT and the solenoid a new Time Of Flight  detector has been 
installed.  The TOF resolution of  order 100 ps allows for 2$\sigma$ $K\pi$ separation for 
transverse momentum up to 1.6 GeV. \par
The calorimeter has been retained from Run I in the
central part, while a new scintillator based plug calorimeter is replacing the old
gaseus calorimeter in the large $\eta$ region. 
It extends to $\eta$ up to 3.6 and maintains
as much as possible the same $\eta\phi$ segmentation of the 
central calorimeter.\par
Finally, the muon system has been partially upgraded: the old Run I 
central muon detectors has been retained but equipped 
with new readout electronics,  while a new extension and intermediate muon chambers will guarantee the muon 
trigger coverage from $|\eta|$ = 0.6 to 1.0. \par

All the front end and DAQ electronics have been changed and upgraded.
At Level 1(within 5$\mu$s)  
a new online processor reconstructing tracks has been implemented 
(eXtremly Fast Tracker) and at Level 2
(within 20 $\mu$s) a Silicon Vertex Trigger (SVT) links the Level 1 
tracks ( from the drift chamber) to
the silicon hits and reconstructs offline-quality tracks 
( with about 40 $\mu$m impact parameter resolution). This is the first time such a device has been
installed in a hadron collider detector and CDF relies upon it to collect
large samples of hadronic $b$ decays  crucial for $B_s^0$ mixing and 
CP violation studies. On the other hand the device has been shown to be very powerful for high
$p_t$ physics, where it allows to select samples enriched in  heavy 
flavors already at the trigger level, that can subsequently be used in 
analysis using heavy flavor tagging.

\subsection{\bf{Search Strategies at CDFII} }

Two different approaches to search for physics beyond the Standard Model are actively pursued in Run II 
in a complementary fashion: model-based analysis and signature based studies.\par
 
In the more traditional model-driven approach, one picks a favorite 
theoretical model and/or a process, and the best signature is chosen. The selection cuts for acceptances are optimized 
based on signal MC. The expected background is calculated from data or Monte Carlo and, 
based on the number of events observed in the data, a discovery is made or the best limit on the new signal is set.
In a signature-based approach a specific signature is picked ( i.e. dijets+X) and the 
data sample is defined in terms of known SM processes. A signal region (blind box) might be defined with 
cuts which are kept as loose as possible and the background predictions in the signal region are often extrapolated
from control regions.  Inconsistencies with the SM predictions will provide indication of possible new physics. 
As the cuts and acceptances are often calculated independently from a model, different models can be tested against the
data sample.\par

In the next sections we will present the first CDF results from both approaches.

\section{\bf{Model based searches}}

\subsection{ \bf{Search for Z' and Randall-Sundrum graviton in high mass dileptons}}

Neutral gauge bosons in addition to the Standard Model Z are expected in many extension of the Standard Model. 
These models
tipically specify the strenghts of the couplings of the Z' to quarks and leptons, but make no prediction for the Z' mass.
A search improving the Run I result has been performed in the di-electrons and dimuons channel, as one of the primary 
observable effects would be anomalous dilepton production at large invariant mass via virtual exchange of the Z'.\par

Anomalous dilepton production at large mass can also be the result of Randall-Sundrum graviton, which would appear as a
resonance of spin 2. Models with extra dimensions have been recently introduced to solve the hierarchy problem\cite{ADD}.
If the extra dimensions are large  the effective Planck scale $M_S$ can be in the TeV range.
In the initial Arkani-Hamed, Dimopoulos, Dvali (ADD) models, gravity can propagate in large extra dimensions and the 
effect of gravity is enhanced at high energy due to the accessibility of numerous excited states of the graviton 
(called Kaluza Klein modes).
In the Randall-Sundrum graviton model\cite{RS} a 4-dimensional metric is multiplied by a {\it{ warp}} factor, 
which is a function of the compactification radius and changes exponentially with the additional dimension. 
Due to the presence of the warp factor, generating a large hierarchy does not require a large curvature radius or 
extra dimension.  The masses and couplings of each individual Kaluza Klein states to matter are determined by 
the warp factor. This implies  that these Kaluza Klein excitations of the graviton can be separately  
produced as resonances, enhancing the DY cross section at large mass.\par
The Drell Yan spectrum in the $ee$ and $\mu\mu$ channel has been compared to the SM expectations 
and no excess has been observed. A 95\% C.L. upper limit on the cross section  as a function of Z' and RS graviton mass.
The Z' limit is better than the Run I measurement in the $ee$ channel ( figure 1-4).

\begin{figure}[ht]
\begin{center}
\begin{minipage}{2.5in}
 \psfig{figure=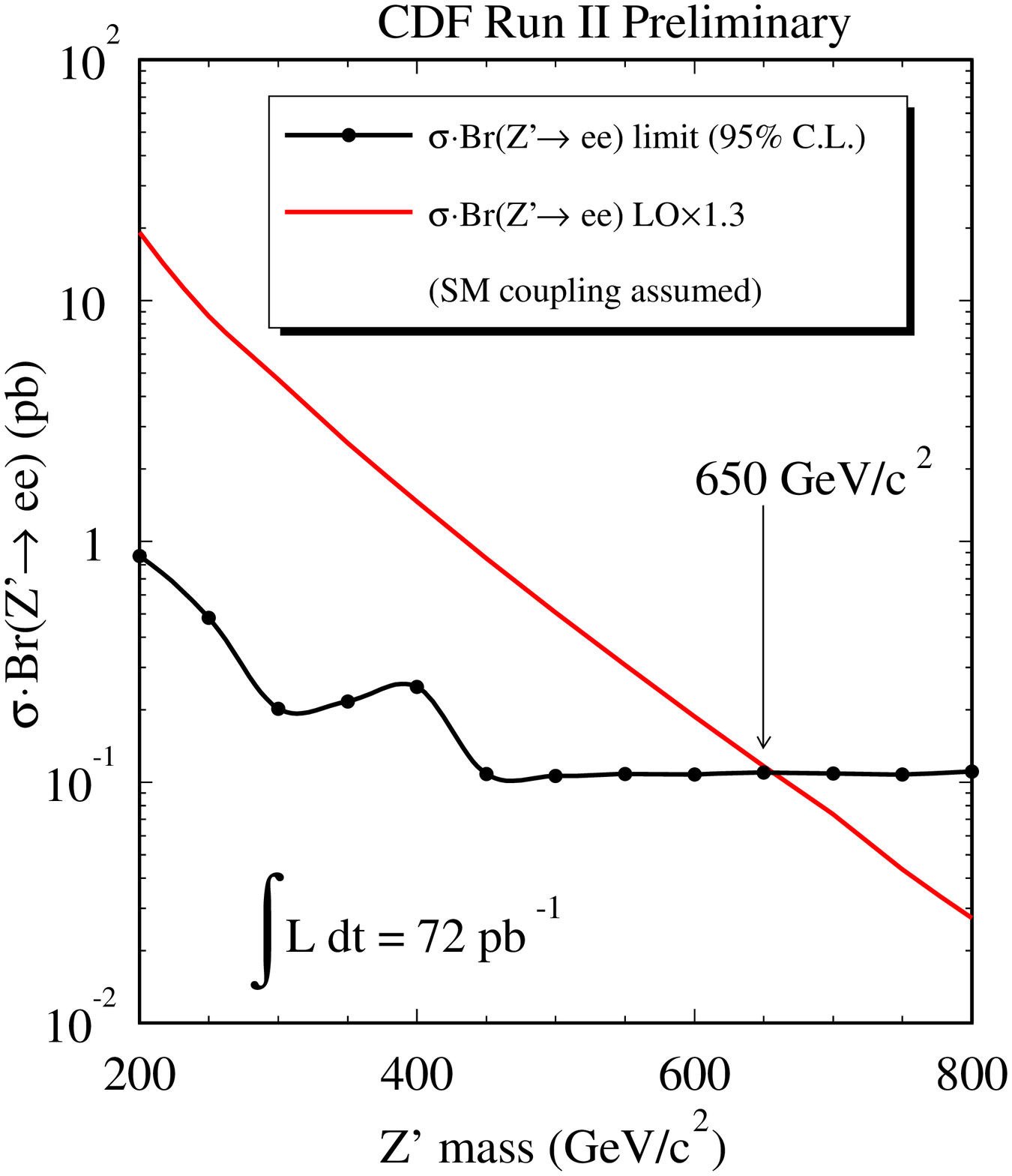,height=1.5in,width=2.5in}
 \caption{Cross section limit as function of the Z' mass, in the dielectron channel}
 \label{fig:Zpee}
\end{minipage}
\hfill
\begin{minipage}{2.5in}
 \psfig{figure=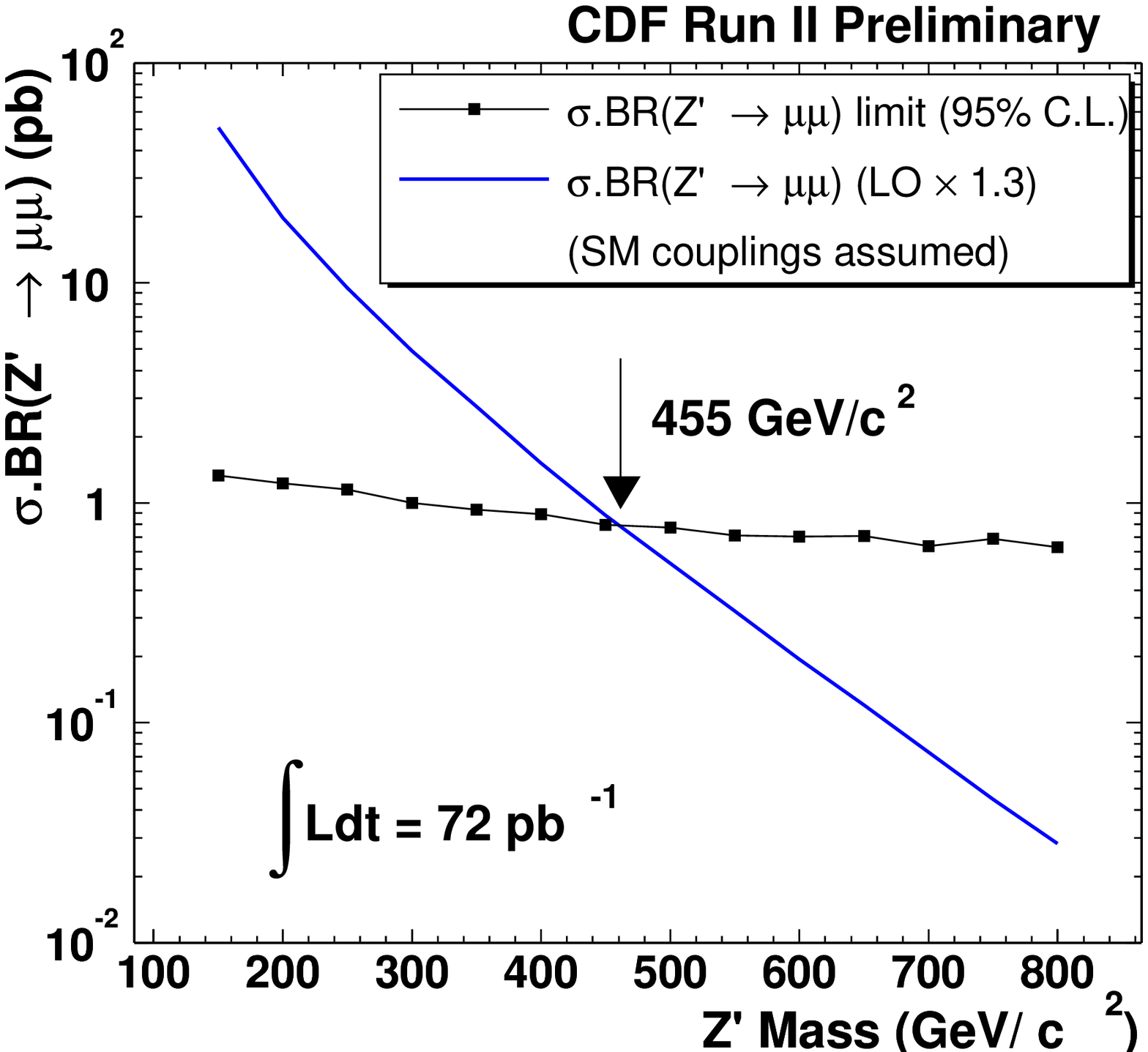,height=1.5in,width=2.5in}
 \caption{Cross section limit as function of the Z' mass, in the dimuon channel}
 \label{fig:Zpmumu}
\end{minipage}
\end{center}
\end{figure}

\begin{figure}[ht]
\begin{center}
\begin{minipage}{2.5in}
 \psfig{figure=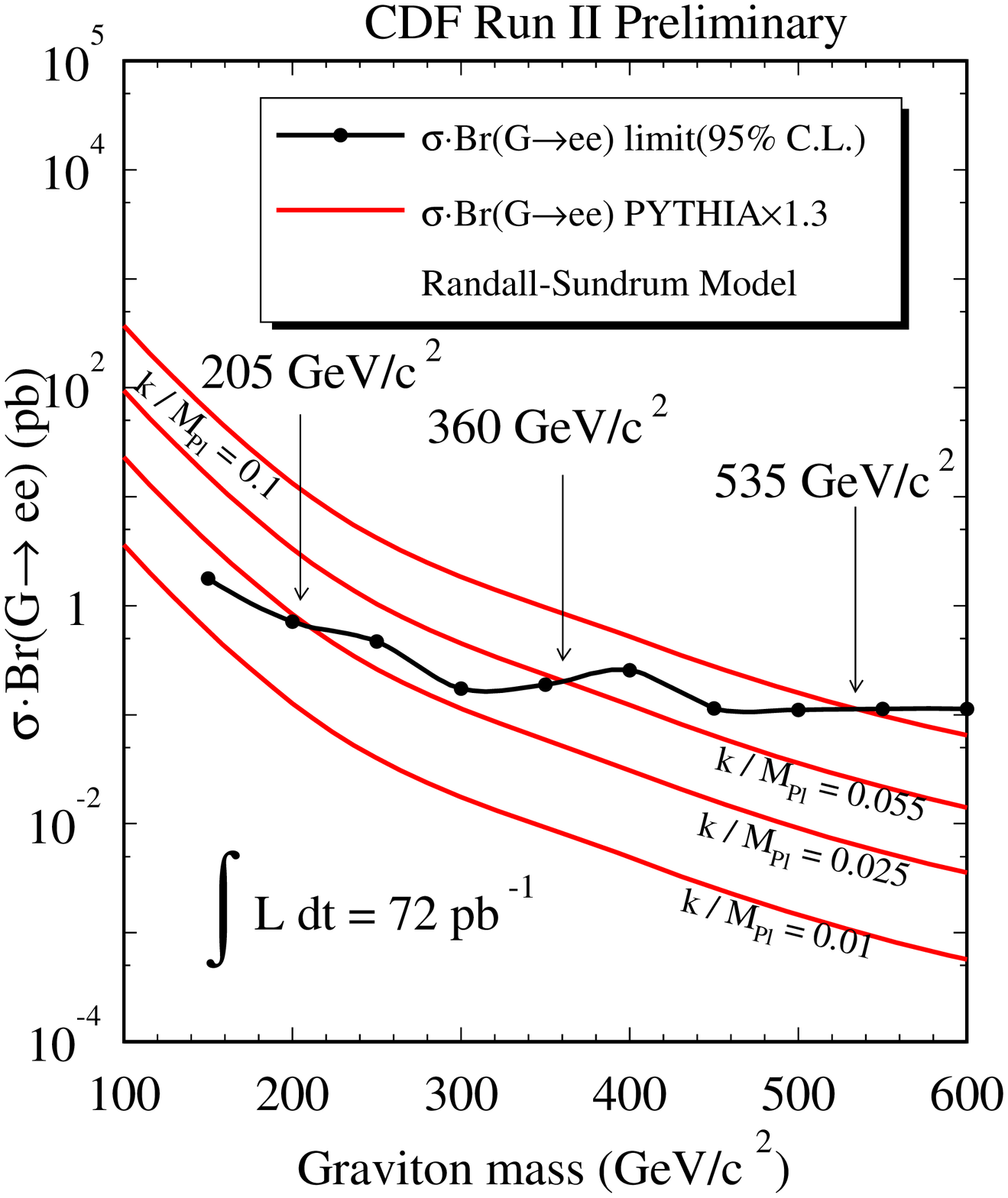,height=1.5in,width=2.5in}
 \caption{Cross section limit as function of the Randall Sundrum graviton mass, in the dielectron channel}
 \label{fig:RSee}
\end{minipage}
\hfill
\begin{minipage}{2.5in}
 \psfig{figure=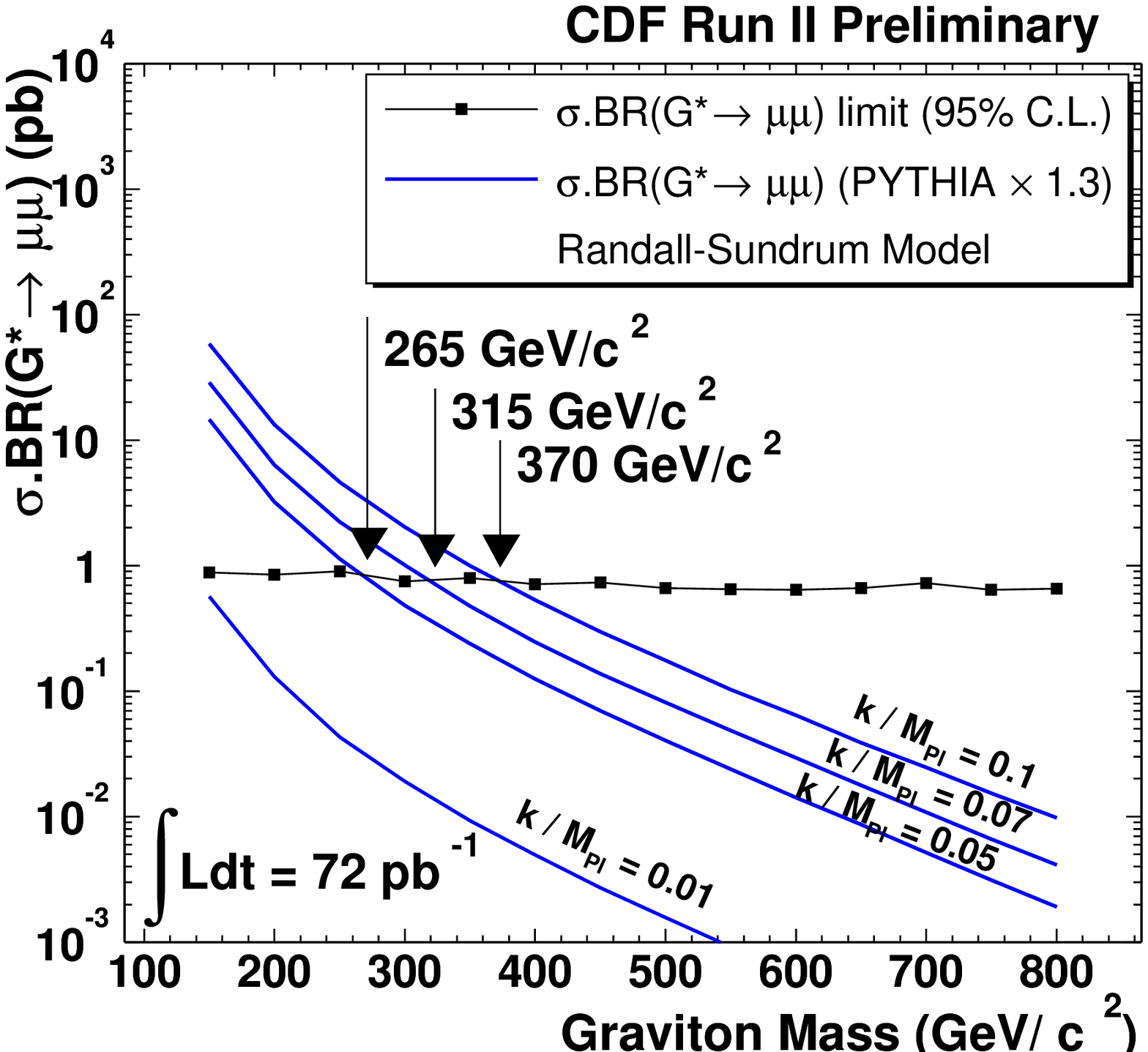,height=1.5in,width=2.5in}
 \caption{Cross section limit as function of the Randall Sundrum graviton mass, in the dimuon channel}
 \label{fig:RSmumu}
\end{minipage}
\end{center}
\end{figure}

\subsection{ \bf{Search for first generation leptoquarks in dielectrons and jets}}

Leptoquarks are hypothetical color-triplet particles carrying both baryon and 
lepton quantum numbers and are predicted by many extension of the Standard 
Model as new bosons coupling to a lepton-quark pair\cite{LQ}. Their masses are not predicted. 
They can be scalar particles (spin 0) or vector (spin 1) and at high energy hadron 
colliders they would be produced directly in pairs, mainly through gluon fusion or quark antiquarks annihilation. 
The couplings of the leptoquarks to the gauge sector are predicted due to the gauge symmetries, 
up to eventual anomalous coupling in the case of vector leptoquarks, whereas the fermionic couplings 
are free parameters of the models. 
In most  models leptoquarks are expected to couple only to fermions of the same generations
because of experimental constraints as non observation of flavor changing neutral currents or 
helicity suppressed decays. 
The NLO cross section at $\sqrt(s)$ = 1960 GeV is about 25\% higher than at 1800 GeV/c$^2$\cite{kramer}.
The current CDF result is focusing on the search for first generation scalar 
leptoquarks, pair produced and decaying both into an electron and a quark. 
The analysis strategy is the following: a reduced data sample is derived from the high $p_T$ inclusive electron
trigger sample. The following cuts are applied:
\begin{itemize}
\item{} 2 electrons with $E_T$ $>$ 25 GeV;
\item{} 2 jets with $E_T$(jet1) $>$ 30 GeV and $E_T$(jet1) $>$ 15 GeV;
\item{} Removal of events lying in the Z mass window (76 $< M_{ee} <$ 110 GeV/c$^2$);
\item{} $E_T$(jet1) + $E_T$(jet2) $>$ 85 GeV and $E_T$(e1) + $E_T$(e2) $>$ 85 GeV;
\item{} $\Sigma((E_T(jet1) + E_T(jet2))^2 +  (E_T(e1) + E_T(e2) )^2 ))> $ 200 GeV;
\end{itemize}
No events survive the analysis cuts.\par
The main background to the process is represented by production $\gamma /Z$ decaying $\to$ $ee$
events accompanied by jets due to radiation. 
The main component of this background is eliminated by cuts on $M_{ee}$ 
around the mass of the Z boson and the $\Sigma E_T$ cuts. 
Another source of background is represented by $t\bar t$ production where both the W decay into e$\nu$. 
Backgrounds from bb, Z$\to \tau\tau$, WW are expected to be negligible due to the an electron isolation cut and 
large electron and jet transverse energy requirements.\par


As no candidate events were found, a 95\% C.L. upper limit on the cross section times a factor $\beta^2$
as a function of m(LQ) has been set, and it is reported in figure 5\footnote{$\beta$ is the branching 
Br(LQ$\to$eq), assumed to be equal to 1 for this analysis.}. A mass limit is set at 230 GeV/c$^2$, better than the 
Run I result.

\begin{figure}[ht]
\begin{center}
  \begin{minipage}{1.8in}
  \psfig{figure=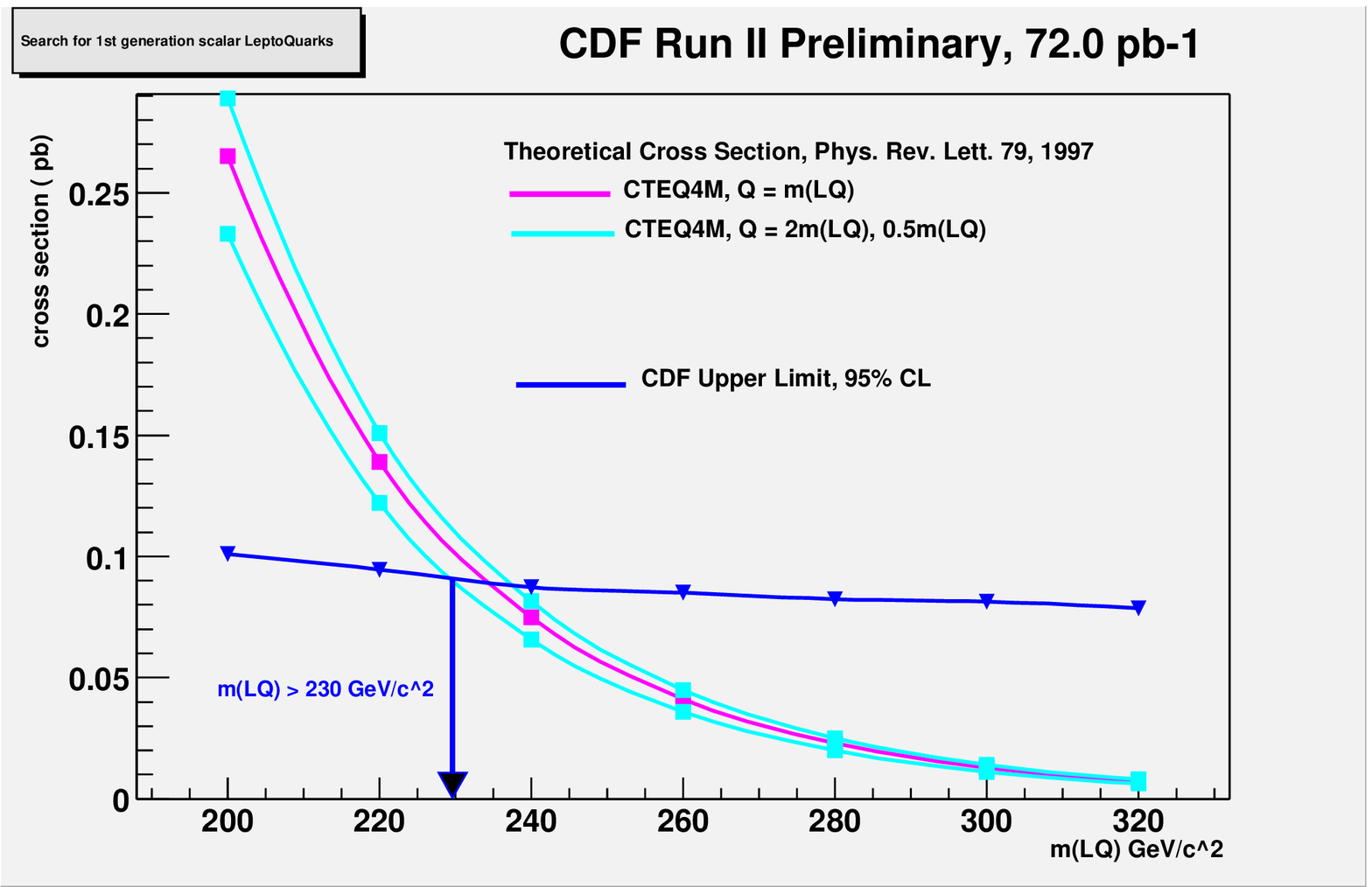,height=1.8in}
  \caption{Limit cross section as a function of m(LQ) compared with the 
     theoretical expectations calculated at NLO accuracy times branching ratio.}
  \label{fig:lq}
  \end{minipage}
\hfill
  \begin{minipage}{1.8in}
  \psfig{figure=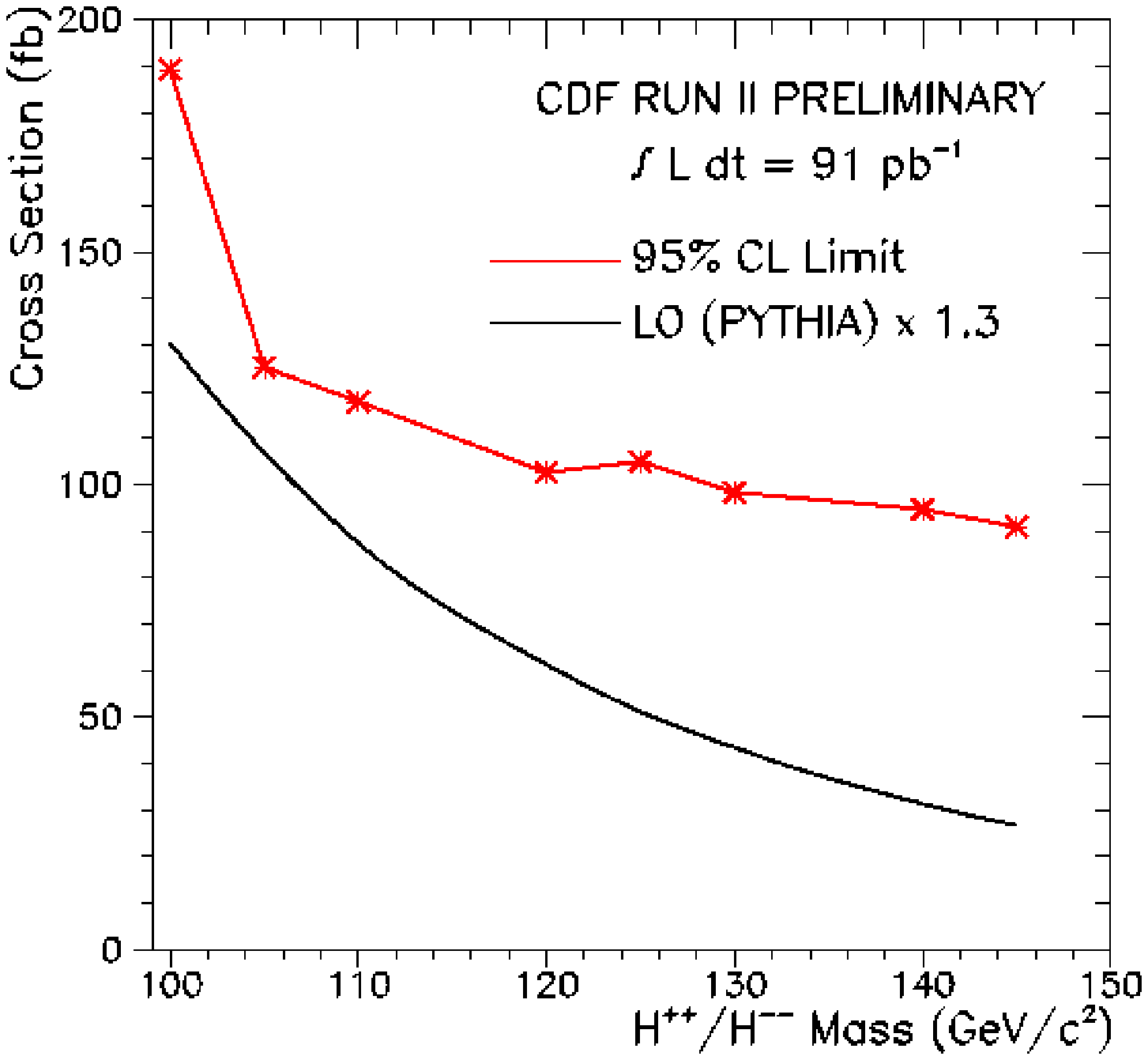,height=1.8in}
  \caption{Limit cross section as a function of m(H++) compared with the 
  theoretical expectations. }
  \label{fig:higgs}
  \end{minipage}
\hfill
\end{center}
\end{figure}

\subsection{ \bf{Search for doubly charged Higgs}}

Doubly-charged Higgs particles are predicted by left-right (LR) symmetric 
models, and SUSY LR models predict low-mass H$^{++}$ (about 100 GeV/c$^2$ to 1 TeV)\cite{Higgs}.  
These  particles decay to leptons, and CDF has searched 91 pb$^-1$ of CDF Run 2 
data for evidence of their production. The search has been performed in  
the same-sign electron mass windows of $\pm$10\% of a given H$^{++}$ mass (about 3 sigma of
the detector resolution). The search is sensitive to any doubly-charged 
particle decaying to dielectrons, whether it is produced singly or in pairs.  
In the 80-100 GeV/c$^2$ mass range, the dielectron search has little 
sensitivity due to the overwhelming background from Z production.  The
background occurs when the Z decays to opposite-sign electrons and one
of the electrons radiates a photon, which subsequently converts.  When the
wrong sign conversion track is associated with the electron cluster, the 
event is reconstructed with 2 same-sign electrons.
The Z background has been evaluated using the data in the 80-100 GeV/c$^2$ mass 
range and the search has been performed in the region above 100 GeV/c$^2$.  
The mass region of 100-130 GeV/c$^2$ is dominated by this background; 
above this mass, QCD and Z processes  are expected to contribute equally to the background

The low mass region ($<$80 GeV/c$^2$) has been used as a test of the background prediction, which is
consistent with 0 events observed.  In the search  region ($>$100 GeV/c$^2$), 0 events are also observed.  
These results provide a 95\%  confidence level cross section limit for pair-production of doubly-charged
particles, which is reported in Figure 6. As the limit cross section sits above the theoretical
prediction it is still not possible to set a limit on the H$^{++}$ mass.

\section{ \bf{Model based searches}}

\subsection{\bf{Search for charged massive particles ( CHAMPS)}}

A search has been performed for charged massive  particles long lived enough to 
escape the CDF detector. As these particles are supposed to
be highly isolated and slow moving, the high $p_T$ muon trigger has been used to select the data sample. 
Particle with $p_T > $ 40 GeV/c are selected to insure full tracking efficiency. 
These particles should also have a  long time of flight through the detector and the new TOF system of CDF 
provides a sensitivity to higher $\beta\gamma$  than a $dE/dx$ measurement (which was used in a similar
search at other machines). The  event $t_0$ is derived from tracks with $p_T$ $>$ 20 GeV/c and it has been tested with  
electrons coming from W decay. Candidate tracks are required to have a high $\Delta_{TOF} = TOF_{tracks} - t_0$.
In figure 7 the predicted number of events is compared to the observed number of events 
in the signal region: $\Delta_{TOF} >$ 2.5 ns.   This point has a background prediction of 
2.9 $\pm$ 0.7 (stat) $\pm$ 3.1 (sys), with 7 events observed.

\begin{figure}[ht]
\begin{center}
  \begin{minipage}{2.2in}
  \psfig{figure=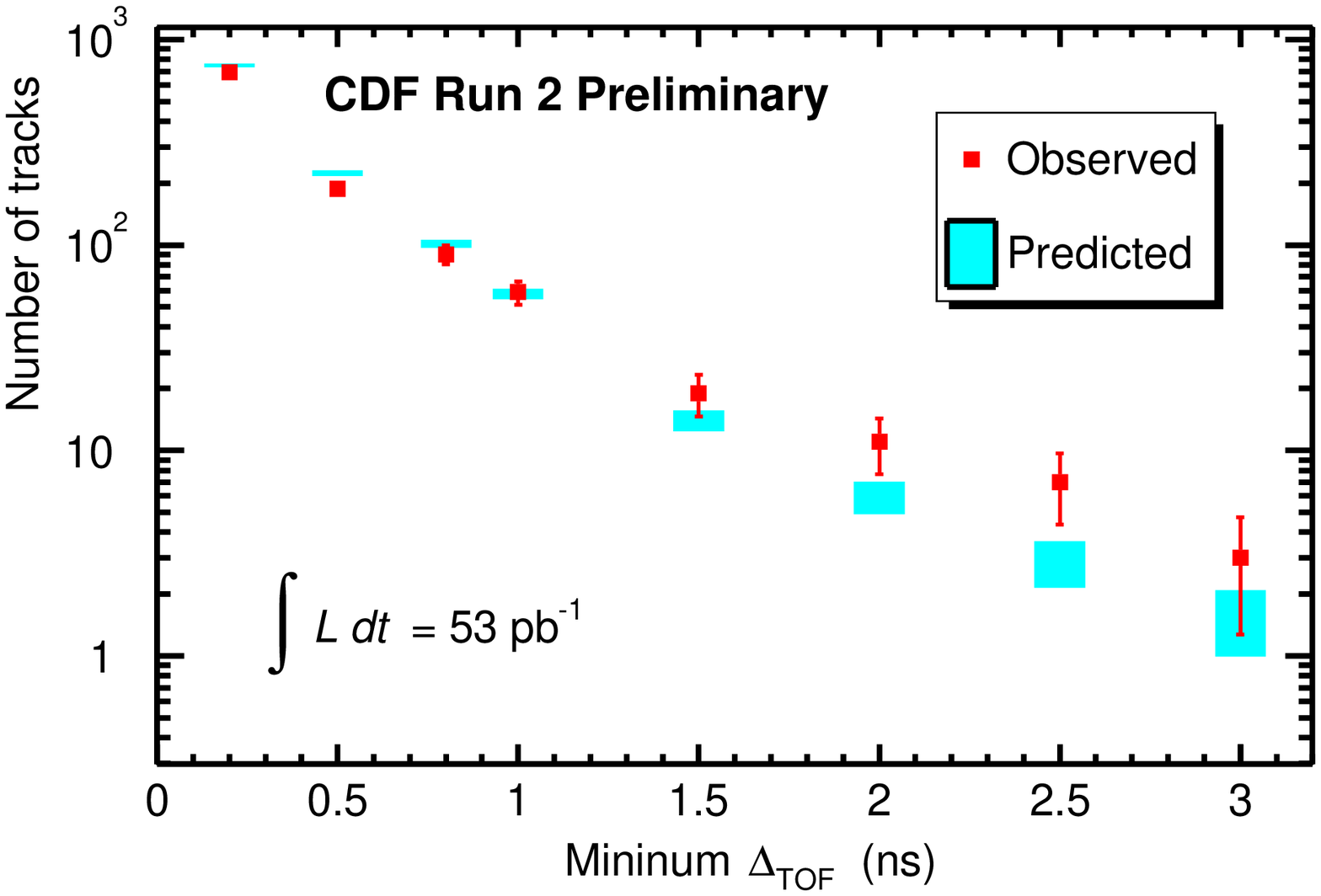,height=1.5in}
  \caption{CHAMPS search: predicted number of events  compared to those observed in the data.}
  \label{fig:champs1}
  \end{minipage}
\hfill
  \begin{minipage}{2.0in}
  \psfig{figure=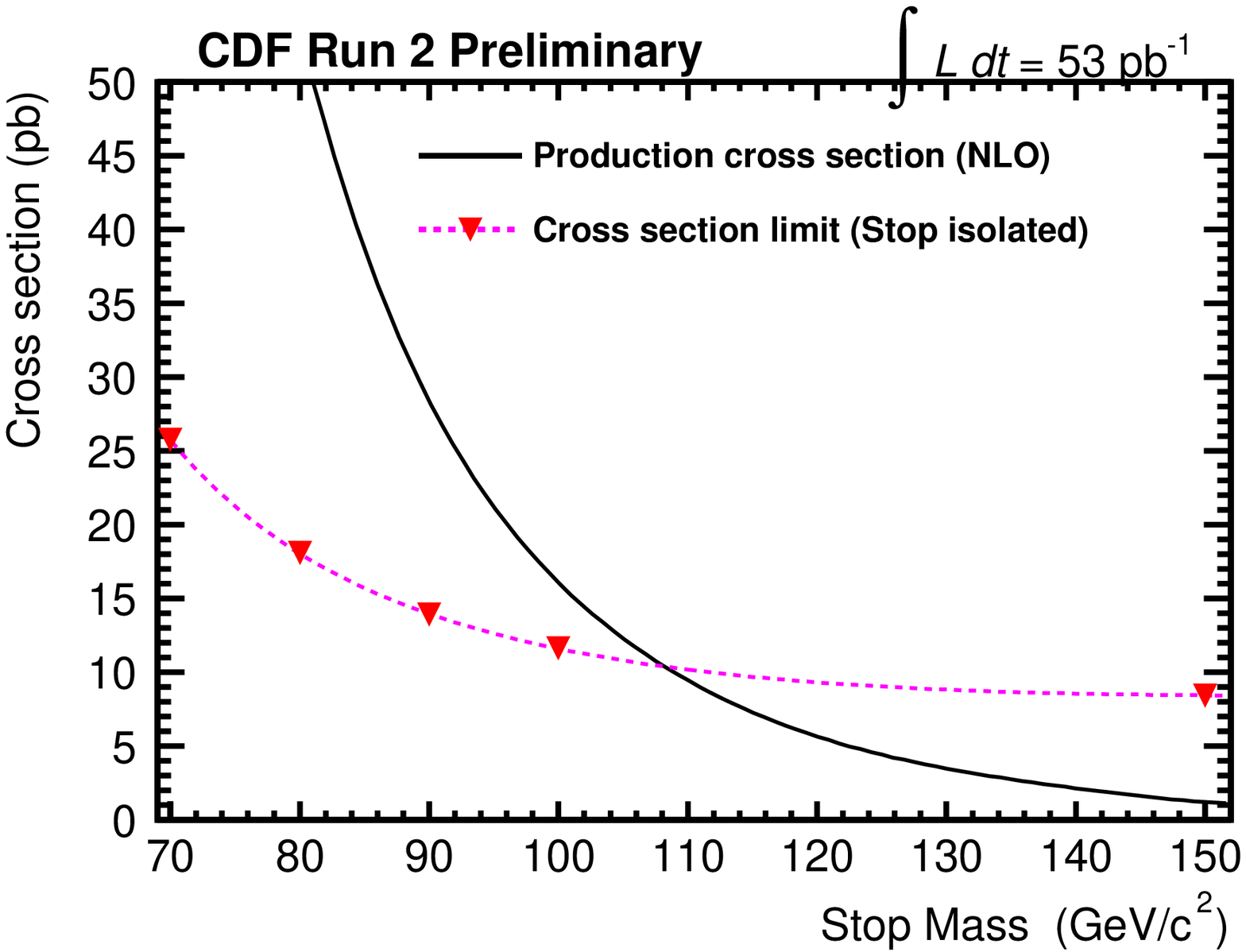,height=1.5in}
  \caption{CHAMPS search: cross section limit as function of stop mass}
\label{fig:champsL} 
  \end{minipage}
\hfill
\end{center}
\end{figure}

Several interesting SUSY-based model can be tested for the presence of slowly moving charged particles. 
In particular, models where  SUSY is broken at low scale, as those including 
gauge-mediated supersymmetry breaking, are generally distinguished by the 
presence of a nearly massless Goldstino as the lightest supersymmetric 
particle and a stop or a stau as the NLSP. 
The next-lightest supersymmetric particle(s) (NLSP) decays to its 
partner and the Goldstino. Depending on the supersymmetry breaking scale, 
these decays can occur promptly or on a scale comparable to or larger 
than the size of a detector ( as for CHAMPS)\cite{champs}.
The stop case has been investigated in this analysis 
and  a 95\% C.L. limit on the cross section for stop production as a 
function of the stop mass has been set.
The resulting mass limit is M(stop) $>$ 108 GeV at 95\% CL and it reported in figure 8.

\subsection {\bf{Search for mass bumps in the dijets spectra}}

A search for new particles decaying to dijets has been performed. The run 2 dijet mass spectrum begins at 
180 GeV and falls steeply.  The highest mass event has a dijet mass of 1364 GeV. 
Comparing the spectrum in Run I and Run II it is evident that  the run 2 cross section 
at $\sqrt(s)$=1.96 TeV  is everywhere above the run 1 spectrum at $\sqrt(s)$=1.80 TeV. 
The ratio of the cross section in Run II to that in Run I compares favorably with a lowest order parton 
level calculation.\par
To search for new particles the mass spectrum has been fitted with a simple background 
parameterization and a search for mass bumps comparable with the CDF mass resolution has been performed. 
The fit, the fractional difference between the data and the fit, 
and the statistical residuals between the data and the fit, show that there is no significant evidence for a 
new particle signal. Several 95\% CL upper limits on the cross section times branching ratio for narrow dijet 
resonances are set and compared with the predictions for axigluons, 
flavor universal colorons, excited quarks, Color Octet Technirhos, E6 diquarks, W' and Z'. 
The limit plot is reported in figure 9.
With the first run 2 data CDF excludes:
\begin{itemize}
\item{} axigluons or flavor universal colorons for masses between 200 and 1130 GeV.\
\item{}excited quarks with mass between 200 and 760 GeV.
\item{} color octet technirhos between 260 and 640 GeV.
\item{} E6 diquarks with mass between 280 and 420 GeV.
\item{} W' with mass between 300 and 410 GeV.
\end{itemize}

All these limits are already better than the corresponding Run I limits.

\subsection {\bf{Search for new physics in the involving photons in the final state}}

Analysis are under way in CDF with the
aim of understanding our datasets in terms of known background contributions and
possible deviation from it as a sign of new physics. 
Signatures involving photons are of particular interest to look for
deviations from the SM predictions in the context of GMSB models.
CDF is also using photon signatures as a first follow-up and check of
anomalous events seen in Run I: in figure 10  the diphotons mass spectra is
reported using approximately 81pb$^{-1}$ of Run II data.

\begin{figure}[ht]
\begin{center}
 \begin{minipage}{2.5in}
 \psfig{figure=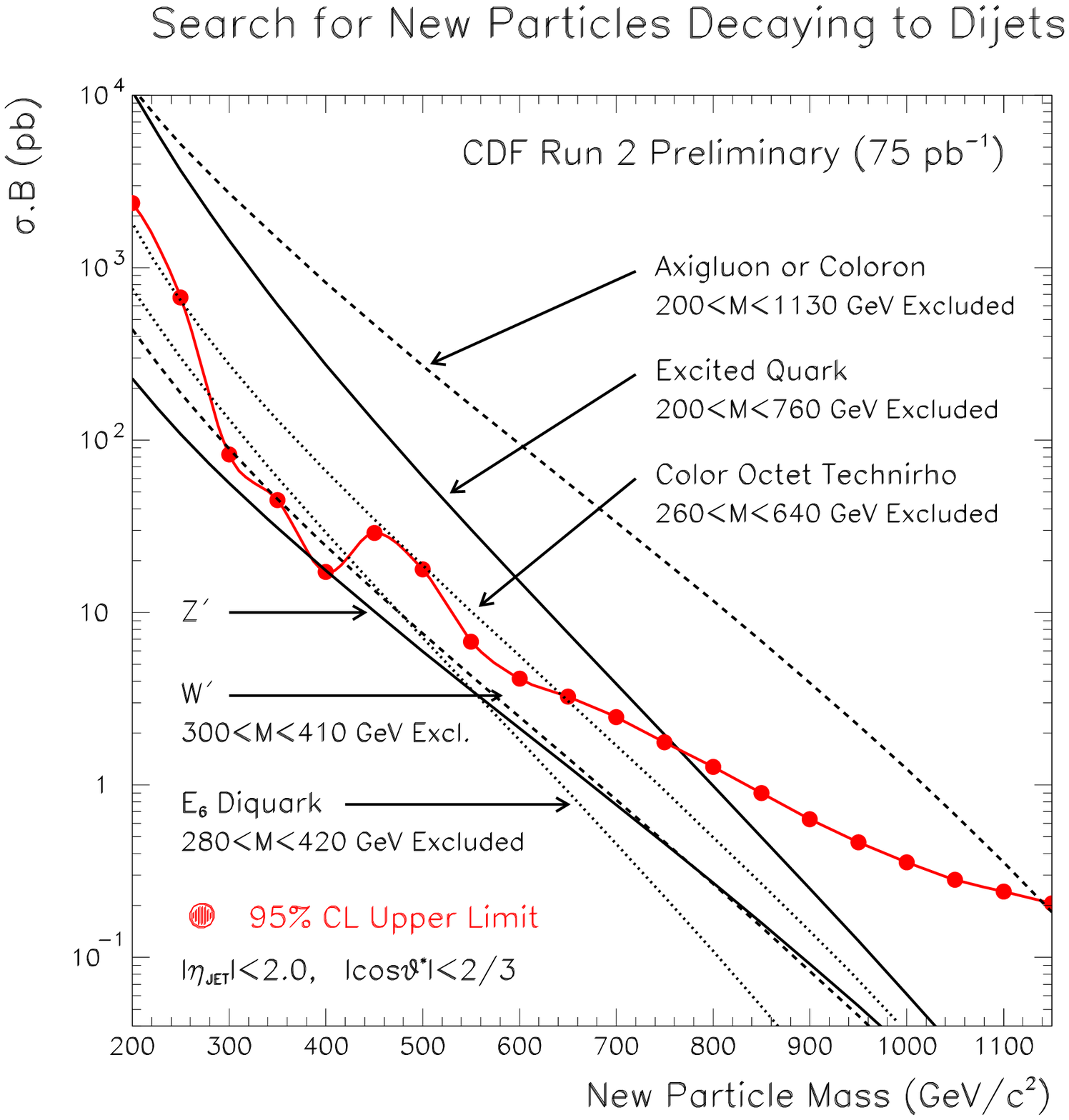,height=1.5in,width=2.2in}
 \caption{95\%CL upper limits on the cross section times branching ratio for 
narrow dijet resonances and compare them with the predictions for axigluons, 
flavor universal colorons, excited quarks, Color Octet Technirhos, E6 diquarks, W' and Z'. }
 \label{fig:dijets}
 \end{minipage}
\hfill
  \begin{minipage}{2.0in}
  \psfig{figure=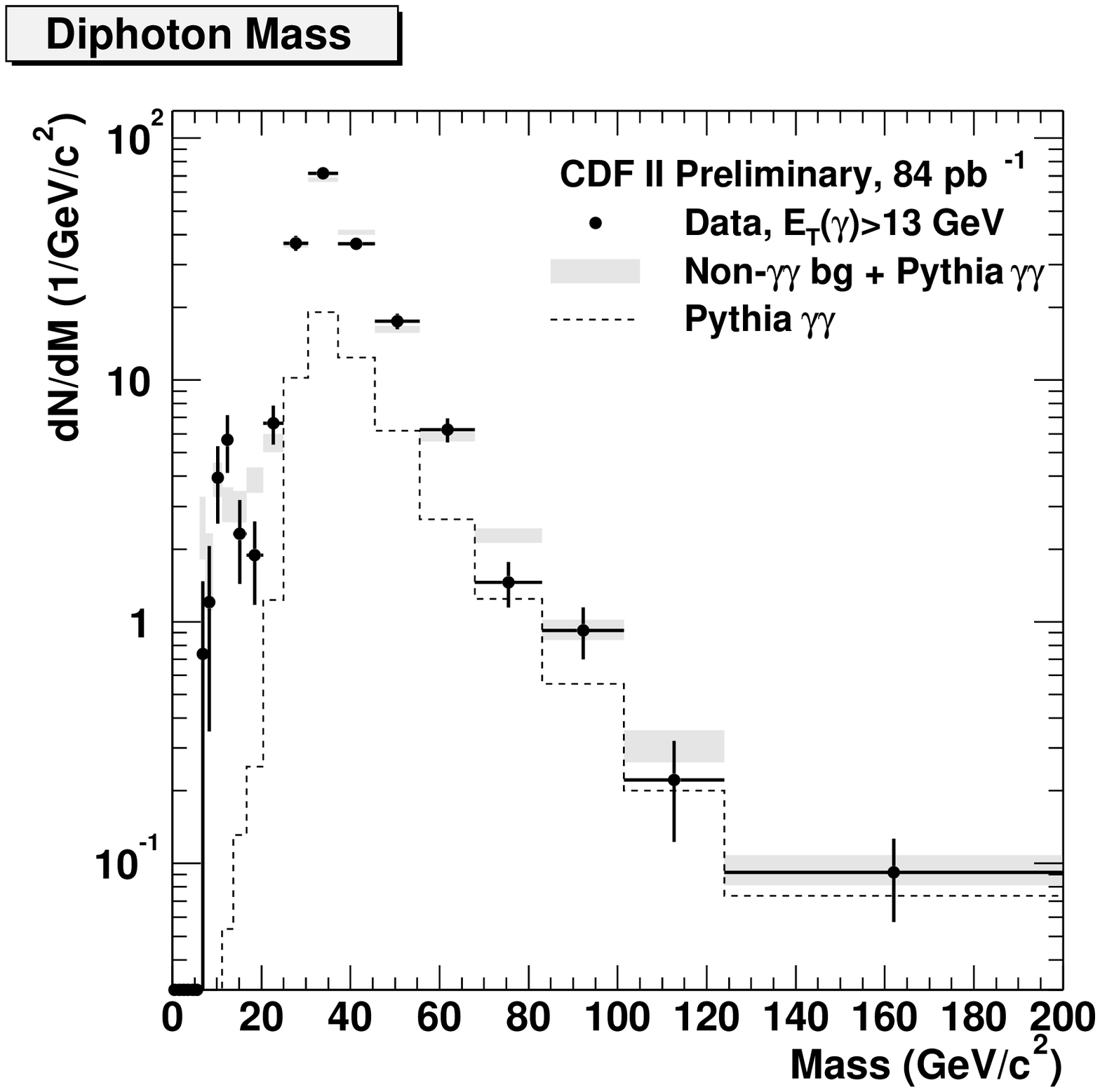,height=2.0in}
  \caption{Diphoton mass spectrum compared to Standard Model expectation}
  \label{fig:photons}
  \end{minipage}
\hfill
\end{center}
\end{figure}

Searches for events with one lepton and diphotons have been carried out and the result ( 0 events observed)
are consistent with SM prediciton.

\section{\bf{Conclusions}}
The CDF experiment is actively taking physics quality data at the upgraded TeVatron collider.
The first results on search for physics beyond the Standard Model have been presented.
In most of the cases they are already competitive if not better than Run I results, 
due to the increase in the center of mass energy and the improved detector capability.

\end{document}